\documentclass{SCGE}
\usepackage{multicol}
\begin{document}

\begin{picture}(0,0){\rm
\put(0,-39){\makebox[160truemm][l]{\bf {\sanhao\raisebox{2pt}{.}}
Research Paper  {\sanhao\raisebox{1.5pt}{.}}}}}
\end{picture}

\def\bm{\boldsymbol}

\def\dl{\displaystyle}
\def\du{\end{document}}
\def\pi{{\uppi}}
\newcommand{\beq}{\begin{equation}}
\newcommand{\eeq}{\end{equation}}
\newcommand{\bea}{\begin{eqnarray}}
\newcommand{\eea}{\end{eqnarray}}

\Year{2014} %
\Month{September}
\Vol{57} %
\No{9} %
\BeginPage{1618} %
\EndPage{1622} %
\AuthorMark{{\rm Dai G F,} et al.}
\DOI{} 


\title{Effect of tensor force on dissipation dynamics in time-dependent Hartree-Fock theory}
\author[1]{DAI Gao-Feng}{}
\author[2,1*]{GUO Lu}{}
\author[1,3]{ZHAO En-Guang}{}
\author[1,3]{ZHOU Shan-Gui}{}

\address[{\rm1}]{State Key Laboratory of Theoretical Physics, Institute of Theoretical Physics, Chinese Academy of Sciences, Beijing 100190, China}
\address[{\rm2}]{School of Physics, University of Chinese Academy of Sciences, Beijing 100049, China}
\address[{\rm3}]{Center of Theoretical Nuclear Physics, National Laboratory of Heavy Ion Accelerator, LanZhou 730000, China}

\maketitle \vspace{-3.5mm}{\footnotesize\begin{center} Received April 25, 2014; accepted ?? ??, 2014
\end{center}}\vspace*{-5mm}

\begin{center}
\rule{16.5cm}{0.4pt}
\parbox{16.5cm}
{\begin{abstract}
The role of tensor force on the collision dynamics of $^{16}$O+$^{16}$O
is investigated in the framework of a fully three-dimensional time-dependent Hartree-Fock theory. The calculations are performed
with modern Skyrme energy functional plus tensor terms.
Particular attention is given on the analysis of dissipation dynamics in heavy-ion collisions.
The energy dissipation is found to decrease as an initial bombarding energy increases in deep-inelastic
collisions for all the Skyrme parameter sets studied here because of the competition between the collective motion and the single-particle degrees of freedom.
 We reveal that the tensor forces may either enhance or reduce the energy dissipation depending on the different parameter sets.
The fusion cross section without tensor force overestimates the experimental value by about 25\%, while the calculation with tensor force T11 has
good agreement with experimental cross section.
\end{abstract}}
\end{center}\vspace*{-0.6cm}
\begin{center}
\parbox{16.5cm}{\bf\jiuhao time-dependent Hartree-Fock, tensor force, dissipation dynamics 
}
\end{center}

\begin{center}
\parbox{16.5cm}{\PACS{\hspace*{-2mm}\rm 24.10.-i, 25.70.-z, 21.60.Jz}
\rule{16.5cm}{0.4pt}}\end{center}



\wuhao\vspace*{1.5mm}
\begin{multicols}{2}
\renewcommand{\baselinestretch}{1.08} \baselineskip 12.2pt\parindent=10.8pt

\no 

\section{Introduction}
The tensor force has been extensively involved in the studies of nuclear structure properties in recent years.
The mean-field calculations have demonstrated the importance of tensor force in the description of ground-state
properties, for example, the spin-orbit splitting~\cite{Colo2007_PLB646-227,Li2013_SCP56-1719} and the appearance of new magic numbers in neutron-rich nuclei~\cite{Marcella2014_PRC89-034316}. The experimental isospin dependence of the
single-particle energies in the chains of $Z=50$ isotopes and $N=82$ isotones has been well accounted for when the tensor interaction is  involved~\cite{Colo2007_PLB646-227}. The Hartree-Fock plus random phase approximation (HF+RPA) calculations have shown that the tensor force has a notable impact on the half-life of
$\beta$ decay~\cite{Minato2013_PRL110-122501} and the magnetic dipole excitation not only due to the changes of the spin-orbit splitting,
but also its RPA correlations~\cite{Cao2009_PRC80-064304}. The Gamow-Teller and charge exchange spin-dipole excitations in $^{90}$Zr and $^{208}$Pb
have also been studied elsewhere~\cite{Bai2009_PRC79-041301,Bai2010_PRL105-072501}. In the framework of
\vspace*{1mm}
\noindent\rule{2.5cm}{0.4pt}\\[0.1mm]{\qihao *Corresponding author (GUO Lu, email: luguo@ucas.ac.cn)}

\no shell model, Otsuka et al.~\cite{Otsuka2001_PRL87-082502,Otsuka2006_PRL97-162501,Otsuka2005_PRL95-232502} have revealed the crucial
role of tensor force in the shell evolution of exotic nuclei.

In spite of these indications in nuclear structure studies, few studies have been performed to explore the role of tensor force in heavy-ion collisions.
Most theoretical models in the description of heavy-ion fusion reactions and deep-inelastic collisions have neglected the tensor component. Because of the complex interplay
between collective motion and single particle degrees of freedom, it is reasonable to expect that some experimental observables should be strongly sensitive to the
tensor force in heavy-ion collisions. The primary purpose of this paper is to investigate the effects of tensor force on dissipation dynamics
and fusion cross section in the framework of a fully microscopic time-dependent Hartree-Fock (TDHF) theory.

TDHF theory has been widely applied to the studies of heavy-ion collisions and resonance dynamics (see reviews elsewhere~\cite{Negele1982_RMP54-913, Simenel2012_EPJA48-152}).
It provides a useful foundation for a fully microscopic many-body theory of large amplitude collective motion. In mean-field dynamics the collective kinetic energy is
converted into the internal degrees of freedom by means of one-body dissipation mechanism. The two-body dissipation, which was included in quantum molecular dynamics models~\cite{Aichelin1991_PR202-233, Wang2002_PRC65-064608, Wen2013_PRL111-012501}, has been neglected within the framework of mean-field theory.
The tensor force has been recently included in TDHF calculations of heavy-ion collisions~\cite{Iwata2011_PRC84-014616} and resonance dynamics~\cite{Fracasso2012_PRC86-044303}. The time-even contribution
of the tensor force was found to become more important in heavy-ion collisions as the increase of the mass of the colliding systems~\cite{Iwata2011_PRC84-014616}.
The full expression of tensor force was suggested by Fracasso et al.~\cite{Fracasso2012_PRC86-044303} to study the resonance dynamics.

Earlier TDHF calculations imposed the various approximations on the effective interaction and geometric symmetry, which caused the systematic underestimation of one-body
energy dissipation. The development of computational power allows a fully three-dimensional TDHF calculation with the modern effective interaction and without symmetry restrictions.
This significantly improved the physical scenario and excited the renewed interests of the studies in heavy-ion collisions~\cite{Umar2006_PRC73-054607, Guo2007_PRC76-014601, Guo2008_PRC77-041301, Washiyama2009_PRC79-024609, Simenel2010_PRL105-192701,
Simenel2011_PRL106-112502, loebl2012_PRC86-024608,Guo2012_EPJWoC38-09003,Simenel2013_PRC88-064604,Sekizawa2013_PRC88-014614,Simenel2014_PRC89-031601}
and resonance dynamics~\cite{Simenel2003_PRC68-024302, Nakatsukasa2005_PRC71-024301, Maruhn2005_PRC71-064328, Umar2005_PRC71-034314, Reinhard2007_EPJA32-19,Simenel2009_PRC80-064309}. The present paper also manages the modern 3D TDHF calculatoins to investigate the effect of tensor force on
the one-body dissipation mechanism. Herein we first briefly review the TDHF theory and the tensor force. Then, we
illustrate the role of tensor force on the dissipation dynamics and fusion cross section, as well as the comparison with experimental data.

\section{Theoretical framework}
\label{theory}
Most TDHF calculations employ the Skyrme effective interaction~\cite{Skyrme1956_PM1-1043}. The Skyrme parameter sets which are widely used in TDHF calculations
do not include the tensor contribution, although the tensor component is present in the original Skyrme force.
The zero-range tensor terms of the Skyrme force can be expressed as thus:
\bea
v_T(\mathbf{r}) &=& \frac{T}{2}\left\{\left[\left(\sigma_{1}\cdot\mathbf{k}'\right)\left(\sigma_{2}\cdot\mathbf{k}'\right)-
\frac{1}{3}\left(\sigma_{1}\cdot\sigma_{2}\right)\mathbf{k}'^{2}\right]\delta\left(\mathbf{r}\right)\right.\nonumber \\
&& \left. \mbox{}+\delta\left(\mathbf{r}\right)\left[\left(\sigma_{1}\cdot\mathbf{k}\right)\left(\sigma_{2}\cdot\mathbf{k}\right)-
\frac{1}{3}\left(\sigma_{1}\cdot\sigma_{2}\right)\mathbf{k}{}^{2}\right]\right\} \nonumber \\
&& \mbox{}+ U\left\{\left(\sigma_{1}\cdot\mathbf{k}'\right)\delta\left(\mathbf{r}\right)\left(\sigma_{2}\cdot\mathbf{k}\right)-
\frac{1}{3}\left(\sigma_{1}\cdot\sigma_{2}\right)\right. \nonumber \\
&& \left. \left(\mathbf{k}'\cdot\delta\left(\mathbf{r}\right)\mathbf{k}\right)\vphantom{\frac{1}{3}}\right\},
\eea
where $\mathbf{r}=\mathbf{r}_1-\mathbf{r}_2$, and the operator $\mathbf{k}=(\nabla_1-\nabla_2)/2i$ acts on the right and $\mathbf{k}'=-(\nabla_1-\nabla_2)/2i$
acts on the left. The coupling constants $T$ and $U$ denote the strength of triplet-even and triplet-odd tensor force, respectively.

The mean-field Hamiltonian includes the effective mass, the spin scalar, spin vector, current, and spin-orbit potentials as shown by Bonche et al.~\cite{Bonche1987_NPA467-115}. The spin-orbit potential is given by
\beq
h_{\rm q}^{\rm s.o.}=i W_q(\mathbf{r})\cdot\left(\mathbf{\sigma}\times\nabla\right),
\eeq
where the index $q$ denotes protons and neutrons, and $\sigma$ is the Pauli matrix.
The tensor force leads to a modification of $W_q(\mathbf{r})$ as:
\beq
W_q(\mathbf{r})=\frac{1}{2}t_4\left(\nabla\rho+\nabla\rho_q\right)+\left(\alpha\mathbf{J}_q+\beta\mathbf{J}_q'\right),
\label{Eq:Wq}
\eeq
where the first term on the right hand side comes from the Skyrme spin-orbit interaction and the second term includes both the central exchange and the tensor
contributions, that is, $\alpha=\alpha_C+\alpha_T$ and $\beta=\beta_C+\beta_T$. The central exchange contributions:
\bea
\alpha_C &=& \frac{1}{8}\left(t_1-t_2\right)-\frac{1}{8}\left(t_1 x_1+t_2 x_2\right),\\
\beta_C &=& -\frac{1}{8}\left(t_1 x_1+t_2 x_2\right),
\eea
are written in terms of the usual Skyrme parameters.
The tensor contributions:
\bea
\alpha_T &=& \frac{5}{12}U,\\
\beta_T &=& \frac{5}{24}(T+U),
\eea
are expressed in terms of the strengths of tensor force.

The parameters of tensor force have been either fitted on top of the existing Skyrme parameter sets or involved by refitting the full set of Skyrme parameters
on the same footing. In order to see the force dependence, we employ the usual Skyrme parameter set plus the tensor force SLy5+T~\cite{Colo2007_PLB646-227},
and a series of new parametrizations T$IJ$~\cite{Lesinski2007_PRC76-014312} to investigate the dissipation dynamics in heavy-ion collisions. These parameters
used herein are shown in Table~\ref{tab:tensorpara}.
Here the Skyrme parameter sets have been fitted with the nuclear ground state properties and no free parameters have been adjusted on the reaction dynamics.
Note that the inclusion of time-even tensor contribution alone breaks Galilean invariance, which will cause the appearance of spurious excitation~\cite{Perlinska2004_PRC69-014316,Maruhn2006_PRC74-027601}.
This was noted in the TDHF calculations when neglecting time-odd spin-orbit interaction~\cite{Maruhn2006_PRC74-027601}.

TDHF equation is expressed as the time evolution of occupied single-particle wave functions:
\beq
i\hbar\frac{\partial\phi_i(t)}{\partial t}=h[\rho(t)]\phi_i(t).
\eeq
The many-body wave functions are approximated as the antisymmetrized independent particle states to assure an exact treatment of Pauli principle during time evolution. Taking the nuclear ground state as an initial wave function, TDHF time evolution is determined by the dynamical unitary propagator.
Note that TDHF theory deal with the nuclear structure and dynamics in a fully consistent manner.

The set of nonlinear TDHF equation is solved on a three-dimensional Cartesian coordinate-space without any symmetry restrictions. We use
the fast Fourier transformation method to calculate the derivatives. The numerical calculations are performed with $n_x$=32,
$n_y$=24, and $n_z$=24 mesh points along the $x$, $y$ and $z$ axis, respectively, for the
light system $^{16}$O+$^{16}$O. The nuclei start at an initial distance $d_0=$16 fm along the $x$ axis. The conservation of total energy
 and particle number is assured during the time evolution by choosing the parameters of grid spacing $\Delta r$=1 fm and time step $\Delta t$=0.2 fm/c.

We also carry out the fusion calculations for the $^{16}$O+$^{16}$O system. For each energy,
the fusion cross section is obtained by the sharp-cutoff formula~\cite{Bonche1978_PRC17-1700} as thus:
\bea
\sigma_{\rm fus} &=& \frac{2\pi}{k^2}\sum_l(2l+1), \nonumber \\
&\approx & \pi (b_{\rm max}^2-b_{\rm min}^2),
\eea
where $k$ is the wave number for the relative motion and the sum is over all even partial waves for which fusion happens.
The quantities $b_{\rm max}$ and $b_{\rm min}$ represent
the maximum and minimum impact parameters at which fusion occurs.

\section{Results and Discussion}
\label{discuss}
The three-dimensional TDHF calculations for heavy-ion collisions of $^{16}$O+$^{16}$O have been performed with modern Skyrme energy functional plus
tensor terms. The present study does not impose any symmetry restrictions in the calculations.
Several groups have studied this system using the TDHF theory on fusion, deep-inelastic collision, and the related dissipation mechanisms ~\cite{Bonche1978_PRC17-1700,Maruhn1985_PRC31-1289,Umar1986_PRL56-2793, Reinhard1988_PRC37-1026, Tohyama2002_PRC65-037601,Maruhn2006_PRC74-027601,Loebl2011_PRC84-034608,Simenel2013_PRC88-024617}.
Recently we clarified the impact of the omitted Skyrme terms and the imposed geometric symmetries in earlier TDHF calculations
on the reaction dynamics for the $^{16}$O+$^{16}$O system~\cite{Dai2014_PRC}. It is demonstrated that the sensitivity of earlier calculations on the Skyrme parametrizations is because of the various approximations on the effective interaction and geometric symmetry. Therefore, the studies of tensor-force effects
based on the modern TDHF calculations
are expected to provide new insight on the dissipation dynamics.

\begin{figure}
\begin{minipage}[t]{0.46\linewidth}
\centering
\includegraphics[width=\textwidth]{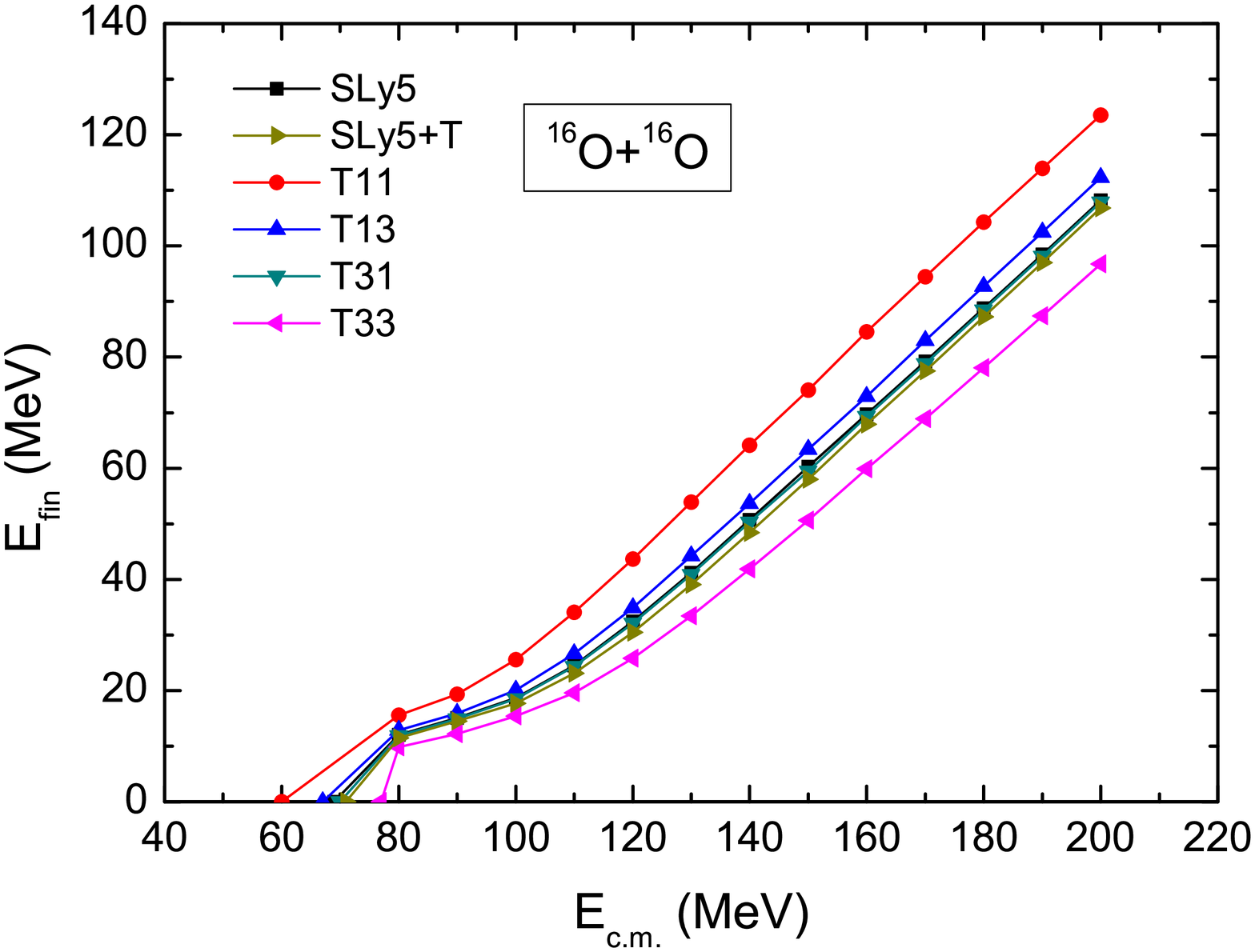}\\
\caption{(Color online) The final relative kinetic energy plotted as a function of initial c.m. energy for head-on collisions of $^{16}$O+$^{16}$O. The lines
  with different colors represent the calculations with the six Skyrme parameter sets.}
\label{Efin}
\end{minipage}
\hfill
\begin{minipage}[t]{0.44\linewidth}
\centering
\includegraphics[width=\textwidth]{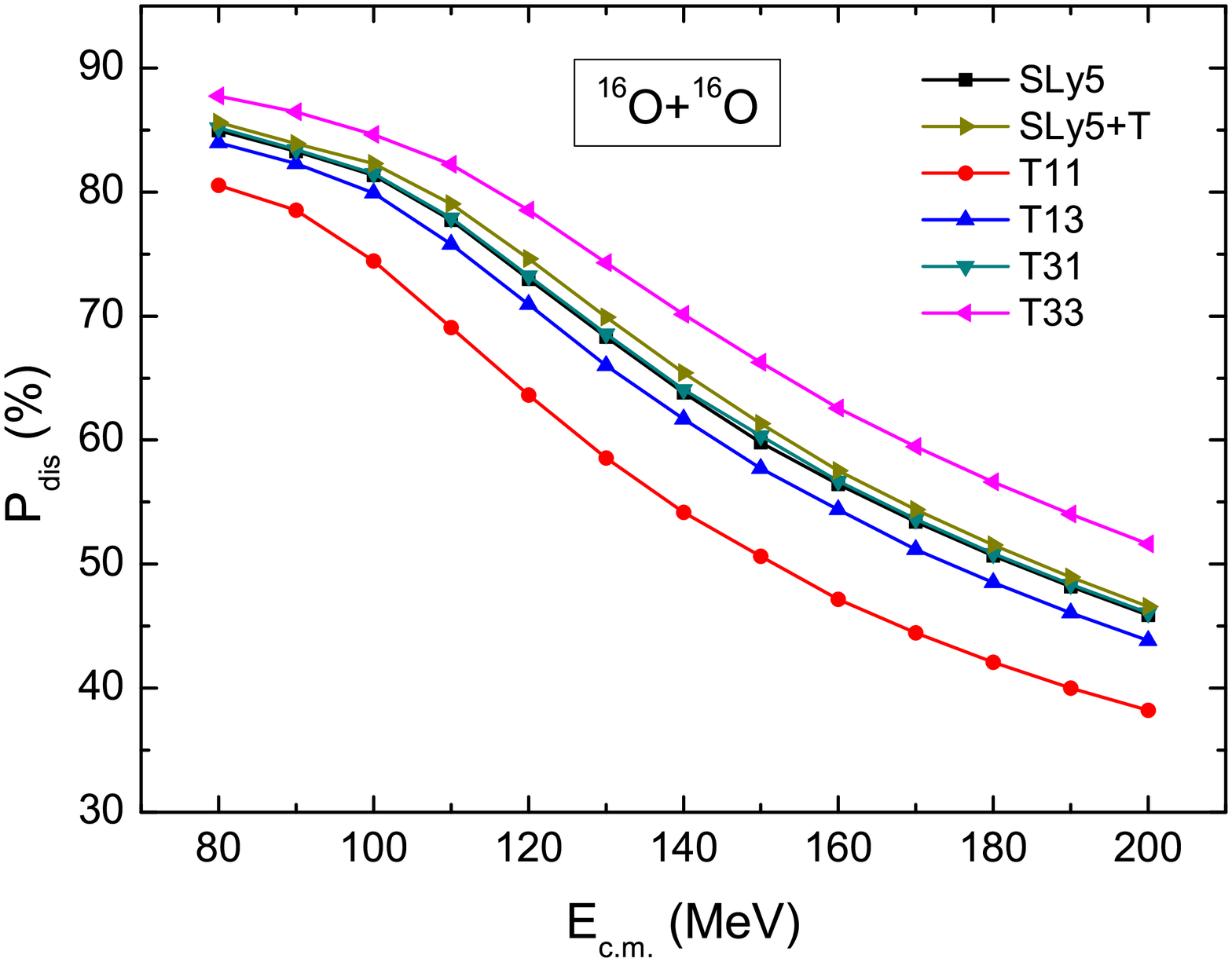}\\
\caption{(Color online) Percentage of energy dissipation as a function of initial c.m. energy for head-on collisions of $^{16}$O+$^{16}$O with the six Skyrme parameter sets.}
\label{Pdis}
\end{minipage}
\end{figure}

One of the most noteworthy experimental observable in heavy-ion scattering is the relative kinetic energy of the separating ions.
Figure~\ref{Efin} shows the final relative kinetic energy of the separating ions as a function of initial center of mass (c.m.) energy for head-on collisions
of $^{16}$O+$^{16}$O. The lines with different colors denote the calculations with six Skyrme parameter sets.
Note that the Skyrme energy functional with SLy5 does not take into account the tensor contribution, while the other five parameter sets in Fig~\ref{Efin}
include the tensor component.
The energy range studied here is from the threshold of inelastic scattering to an initial c.m. energy of 200 MeV.
From Fig.~\ref{Efin}, we can observe that the final kinetic energy increases as the initial energy for all the six Skyrme parameter sets, and yet
there gradually appears distinct discrepancy among the final kinetic energies as the increase of initial energy. The calculations with parameter set T11 give rise to the
largest final kinetic energy and T33 with the smallest energy. It may be expected that the energy dissipation with T33 will be the most notable, and the
most weak with T11 among the six parameter sets. Because the strong dissipation leads to the higher threshold of inelastic scattering~\cite{Umar1986_PRL56-2793},
the threshold energy would be lowest for T11 and highest for T33 as shown in Fig~\ref{Efin}.

In order to illustrate the energy dependence of dissipation dynamics, a quantity which measures the extent of
 energy dissipation in deep-inelastic collisions is defined as $P_{\rm dis}=1-E_{\rm fin}/E_{\rm c.m.}$, where $E_{\rm fin}$ and
${E_{\rm c.m.}}$ denote the final and initial kinetic energy, respectively. Figure~\ref{Pdis} plots the percentage of energy dissipation as a function of initial c.m. energy for head-on collisions of $^{16}$O+$^{16}$O using the six Skyrme parameter sets.
For all the cases the energy dissipation decreases as the incident energy increases. This behavior of energy dissipation, as clarified by Dai et al.~\cite{Dai2014_PRC}, is
because of the competition of collective motion and single-particle degrees of freedom.
By comparing the results without tensor force (SLy5) and with tensor (the other five parameter sets), we found that the tensor contribution may either
increase or decrease the energy dissipation in heavy-ion collisions depending on the parameters of tensor force. Since the spin-orbit
density $\mathbf{J}_q$ and $\mathbf{J}_q'$ shown in Eq.~(\ref{Eq:Wq}) are similar for collisions between $N=Z$ nuclei,
the force parameter dependence mostly arises from the sum of $\alpha$ and $\beta$. With the small value of $|\alpha+\beta|$, the small contribution
to the spin-orbit potential in Eq.~(\ref{Eq:Wq}) results in the small effect of tensor force on the energy dissipation, as the cases of SLy5+T, T13 and T31.
Conversely, the large value of $|\alpha+\beta|$ gives rise to the strong effect on the energy dissipation as T11 and T33. From the comparison of the results with SLy5 and SLy5+T, the tensor force with negative $\alpha+\beta$ enhances the energy dissipation. This is consistent with the analysis of entrance channel dynamics as suggested by Iwata and Maruhn~\cite{Iwata2011_PRC84-014616}.
However, for the T$IJ$ parameter sets it is hard to compare the contributions arising from the tensor force because both the usual Skyrme and tensor parameters have been fitted by using a full variational procedure. We also observe that the energy dissipation is quite sensitive
 to the value of $\beta$. The larger value of $\beta$ results in the larger energy dissipation. For the case with the same
 value of $\beta$, the larger value of $\alpha$ causes the larger dissipation.

The fusion calculations for the $^{16}$O+$^{16}$O system have been carried out at a c.m. energy of 70.5, where the experimental data is available. In light systems the reaction cross section is approximately equal to fusion cross section
due to the high fission barrier, therefore the TDHF results and experimental data can be compared.
At high collision energy, there usually appears central transparency predicted in TDHF calculations and the lower limit of orbital angular momentum will be nonzero. Hence,
both the maximum and minimum impact parameter $b_{\rm max}$ and $b_{\rm min}$ at which fusion occurs are searched within an interval of 0.1 fm.
Table~\ref{tab:fusion} shows the fusion cross section using the six Skyrme parameter sets at a c.m. energy of 70.5 MeV, and the experimental reaction cross section with errors~\cite{SaintLaurent1979_327-517}.
The fusion cross section without the tensor force (SLy5) overestimates the experimental value by about 25\%. Note that TDHF has no free parameters adjusted for the reaction dynamics. Since the strong dissipation results in an enhanced cross section, as we expected,
the cross section is enhanced for SLy5+T, T31 and T33 and reduced for T11 and T13 when the tensor correlation has been taken into account. For T11 a small fusion window with $b_{\rm min}$=2.3 fm and $b_{\rm max}$=6.5 fm results in the smallest fusion cross section.
The good agreement between the result with parameter set T11 and experimental data suggests that T11 might be a good choice among the large number of tensor parameter sets to study the heavy-ion collision dynamics.


\section{Conclusions}
\label{summary}
We have investigated the effect of tensor force on the dissipation dynamics and fusion cross sections in heavy-ion collisions of $^{16}$O+$^{16}$O within the framework of the microscopic
TDHF theory. The calculations have been implemented in a fully three-dimensional coordinate space with modern Skyrme energy functional plus tensor terms.
The dissipation dynamics exhibits a universal behavior for all the Skyrme parameter sets studied here.
We revealed that
the energy dissipation in deep-inelastic collisions of the light systems decreases as an initial  bombarding energy increases because of the interplay between the
collective motion and the single particle motion.
The energy dissipation is sensitive to the parameters of tensor force. The large value of $|\alpha+\beta|$ results in the strong effect of tensor force on the energy dissipation. The tensor force may either increase or decrease the energy dissipation depending on the different parameter sets.
We also performed the fusion calculations for $^{16}$O+$^{16}$O at a c.m. energy of 70.5 MeV. The cross section without the tensor force using SLy5
overestimated the experimental value by about 25\%, while the calculations with tensor force T11 had good agreement with experiment.
We can conclude that the tensor force is crucial in heavy-ion collisions of light systems with respect to the energy dissipation and fusion cross section.

\Acknowledgements{\bahao This work was supported by Natural Science Foundation of China (Grants Nos. 11175252, 11121403, 11120101005, 11211120152, and 11275248),
the National Key Basic Research Program of China (Grant No. 2013CB834400), the Knowledge Innovation Project of the Chinese Academy of Sciences (Grant No. KJCX2-EW-N01), the President Fund of UCAS,
the Scientific Research Foundation for the Returned Overseas Chinese Scholars, Ministry of Education of China, and
the Open Project Program of State Key Laboratory of Theoretical Physics,
Institute of Theoretical Physics, Chinese Academy of Sciences, China (No.Y4KF041CJ1).
The results described in this paper were obtained on the High-performance Computing Clusters of ITP/CAS and the ScGrid of the Supercomputing Center, Computer Network Information Center of the Chinese Academy of Sciences.}


\normalsize \vskip0.3in\parskip=0mm \baselineskip 18pt
\renewcommand{\baselinestretch}{1.1}\footnotesize\parindent=4mm\bahao

\bibliographystyle{sci-chin}
\bibliography{ref}
\end{multicols}
\clearpage

\begin{table}
\begin{minipage}[t]{0.5\linewidth}
\centering
\caption{The parameters of central exchange and tensor contributions as well as the $\alpha$ and $\beta$ values. All values are in MeV $\cdot$ $\rm fm^{-5}$.}
\label{tab:tensorpara}
\begin{tabular*}{\textwidth}{@{\extracolsep{\fill}}lcccccc}
\hline
\hline
Force  &  $\alpha_\mathrm T$ & $\beta_\mathrm T$ & $\alpha_\mathrm C$ &  $\beta_\mathrm C$ & $\alpha$ &	 $\beta$  \\
\hline
SLy5       &     0.0    &	   0.0   &	80.2   & $-$48.9  &    80.2 &  $-$48.9  \\
SLy5+T     &  $-$170.0  &    100.0   &   80.2  & $-$48.9  & $-$89.8 &     51.1  \\
T11        &  $-$142.8  &	 $-$17.5 &   82.8  & $-$42.5  & $-$60.0 &  $-$60.0  \\
T13        &  $-$21.6   &    $-$15.1 &   81.6  & $-$44.9  &    60.0 &  $-$60.0  \\
T31        &  $-$159.4  &    74.2	 &   99.4  & $-$14.2  & $-$60.0 &     60.0  \\
T33        &  $-$40.8   &    71.1    &  100.8  & $-$11.1  &    60.0 &     60.0  \\
\hline
\hline
\end{tabular*}
\end{minipage}
\hfill
\begin{minipage}[t]{0.46\linewidth}
\centering
\caption{Calculations of fusion cross section for $^{16}$O+$^{16}$O at $E_{\rm{c.m.}}=70.5$ MeV using six Skyrme parameter sets, and
the experimental data with errors~\cite{SaintLaurent1979_327-517}.}
\label{tab:fusion}
\begin{tabular*}{\textwidth}{@{\extracolsep{\fill}}lr}
\hline
\hline
Force  &  $\sigma_{\rm{fus}}$ (mb)      \\
\hline
SLy5   &    1307           \\
SLy5+T &    1327            \\
T11    &    1161            \\
T13    &    1265         \\
T31    &    1326         \\
T33    &    1327          \\
Expt.   &   $1056\pm125$    \\
\hline
\hline
\end{tabular*}
\end{minipage}
\end{table}

\end{document}